\documentclass[sigconf]{acmart}

\AtBeginDocument{%
  }

\setcopyright{acmlicensed}

\copyrightyear{2026}
\acmYear{2026}
\setcopyright{cc}
\setcctype{by}
\acmConference[KDD '26]{Proceedings of the 32nd ACM SIGKDD Conference on Knowledge Discovery and Data Mining V.2}{August 09--13, 2026}{Jeju Island, Republic of Korea}
\acmBooktitle{Proceedings of the 32nd ACM SIGKDD Conference on Knowledge Discovery and Data Mining V.2 (KDD '26), August 09--13, 2026, Jeju Island, Republic of Korea}
\acmDOI{10.1145/3770855.3818503}
\acmISBN{979-8-4007-2259-2/2026/08}

\usepackage{makecell}
\usepackage{multirow}
\usepackage[ruled,linesnumbered]{algorithm2e}
\usepackage{booktabs}   
\usepackage[table]{xcolor} 

\usepackage{microtype}
\newcommand{\ie}{\emph{i.e., }}
\newcommand{\eg}{\emph{e.g., }}

\usepackage{xcolor} 

\usepackage{hhline}
\newcommand{\hgreen}[1]{\textcolor{green!60!black}{#1}}
\usepackage{balance}

\begin{document}

\title{UniNote: A Unified Embedding Model for Multimodal Representation and Ranking}

\author{Jinghan Zhao}
\email{zhaojinghan2003@gmail.com}
\authornote{Work done during internship at Xiaohongshu.}
\affiliation{%
  \institution{Xiaohongshu}
  \city{Beijing}
  \country{China}
}

\author{Wenwei Jin}
\email{wenwei1217.jin@gmail.com}
\authornote{Corresponding author.}
\affiliation{%
  \institution{Xiaohongshu}
  \city{Beijing}
  \country{China}
}

\author{Anqi Li}
\email{anqi.li@sjtu.edu.cn}
\affiliation{%
  \institution{Shanghai Jiao Tong University}
  \city{Shanghai}
  \country{China}
}

\author{Jintao Tong}
\email{jintaotong@hust.edu.cn}
\affiliation{%
  \institution{Huazhong University of Science and Technology}
  \city{Wuhan}
  \state{Hubei}
  \country{China}
}

\author{Luya Mo}
\email{3120230882@bit.edu.cn}
\affiliation{%
  \institution{Beijing Institute of Technology}
  \city{Beijing}
  \country{China}
}

\author{Jiawei Li}
\email{wangdesheng@xiaohongshu.com}
\affiliation{%
  \institution{Xiaohongshu}
  \city{Beijing}
  \country{China}
}

\author{Bin Li}
\email{libin656712945@gmail.com}
\affiliation{%
  \institution{Xiaohongshu}
  \city{Beijing}
  \country{China}
}

\author{Yao Hu}
\email{yaoohu@gmail.com}
\affiliation{%
  \institution{Xiaohongshu}
  \city{Beijing}
  \country{China}
}

\renewcommand{\shortauthors}{Jinghan Zhao et al.}

\begin{abstract}
Item-to-Item (I2I) retrieval is a fundamental part of modern content platforms, supporting critical industrial workflows from recommendation engines to content auditing. While multimodal embedding methods have advanced general retrieval, they often falter in I2I scenarios due to the challenges of balancing global content representation with fine-grained local retrieval, the systemic inefficiency of decoupled embedding-and-ranking pipelines, and the inherent trade-offs between model precision and serving latency. To solve these issues, we propose \textbf{UniNote}, a unified embedding model designed for industrial I2I retrieval. Tailored retrieval strategies are introduced to support representation learning over complex, multimodal content at varying granularities. To operationalize these strategies, UniNote employs a two-stage training paradigm: the first stage leverages contrastive SFT to establish robust base embeddings, while the second stage refines ranking quality through a reinforcement learning (RL) process that aligns the model with content relevance. Our results show that UniNote achieves SOTA performance across diverse I2I tasks. Deployed at Xiaohongshu and integrated with Matryoshka Representation Learning (MRL), UniNote achieved significant improvements in retrieval quality and cost efficiency in large-scale applications.






\end{abstract}






\maketitle

\section{Introduction}

\begin{figure*}[t]
  \includegraphics[width=\textwidth]{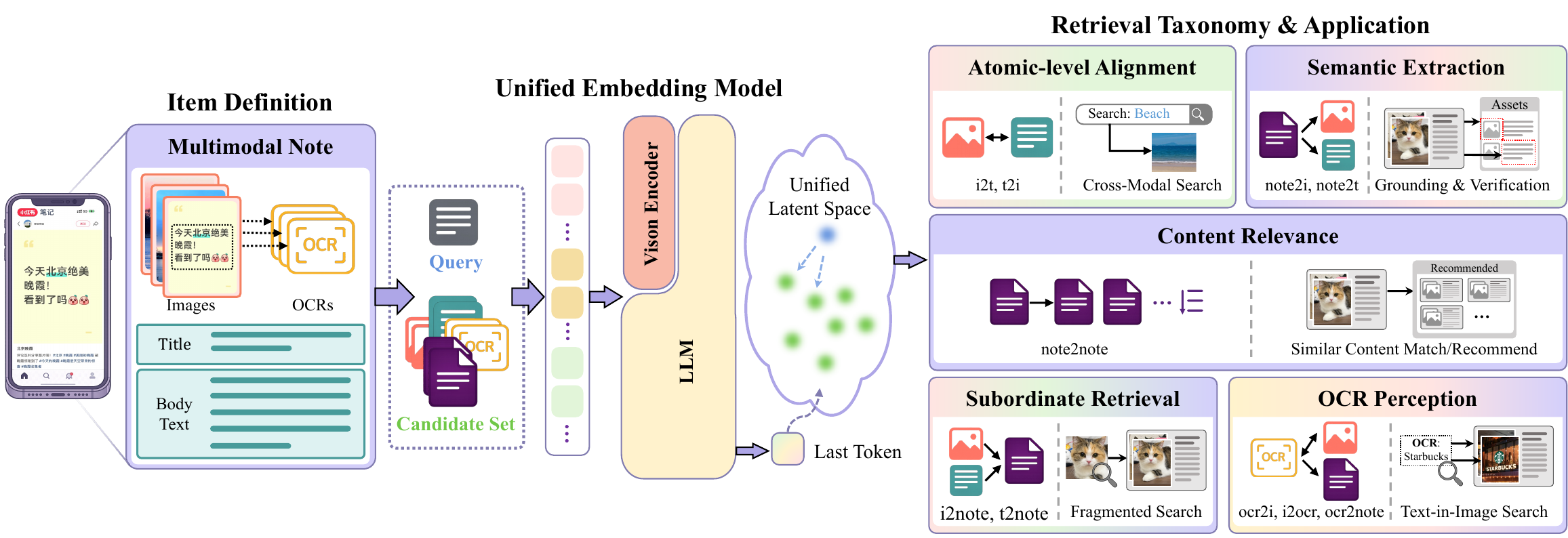}
  \caption{ We propose UniNote, a unified retrieval-ranking framework specifically designed for I2I retrieval, which is capable of encoding various modality inputs into a unified space with cross-modal alignment (task: Atomic-level Alignment, OCR Perception), localized retrieval (task: Semantic Extraction, Subordinate Retrieval), and relevance-aware ranking (task: Content Relevance) capabilities.}
  \label{fig:teaser}
\end{figure*}


I2I retrieval constitutes a fundamental paradigm powering large-scale ecosystems in modern internet content platforms. As content modalities diversify and information volumes surge, retrieval requirements have evolved to be more multifaceted and demanding. Contemporary retrieval frameworks must adapt to heterogeneous inputs across various modalities, including text, images, videos, and their complex combinations, to support a broad range of applications such as recommendation systems\cite{liu2024multimodal,wei2019mmgcn,wu2022mm,xu2024large}, content moderation\cite{yuan2024rethinking,li2025towards,arya2024mscmgtb}, and data management.

User-generated \textbf{notes} on the Xiaohongshu platform serve as a representative example of complex items that typically encompass heterogeneous modalities including text, images, videos. The retrieval of such notes entails several critical requirements. Effective retrieval of these notes imposes distinct functional requirements. First, the system must integrate multimodal information of each note into a cohesive global representation. Second, it necessitates fine-grained local retrieval capabilities, enabling tasks such as locating a full note based solely on an individual image. Third, the architecture is required to incorporate accurate relevance ranking while strictly adhering to the latency constraints of massive industrial databases.

Despite the progress in multimodal embedding methods, addressing these requirements remains challenging due to three fundamental limitations in existing paradigms. 
(1) Structural disconnect in dual-tower models. Standard retrieval architectures (\eg CLIP\cite{radford2021learning} and its variants\cite{zhai2023sigmoid}\cite{li2022blip}) encode modalities independently to ensure efficiency. However, this late-interaction mechanism inherently fails to capture the fine-grained alignment between text and visual components required for composite notes, leading to a loss of local detail. (2) Granularity mismatch in generative baselines. While multimodal large language model (MLLM)-based embeddings\cite{lee2024nv}\cite{zhang2024gme} excel at global reasoning, they suffer from a granularity mismatch. These models are optimized for broad semantic descriptions or question-answering tasks, often overlooking the hierarchical structure needed to retrieve a whole note based on a partial input (\ie local-to-global retrieval). 
(3) Inefficiency in relevance ranking. Regarding ranking, current industrial pipelines typically rely on a disjointed "retrieve-then-rerank" strategy\cite{gu2025unime}\cite{li2026qwen3}. They use heavy generative models to rerank candidates, creating a significant computational bottleneck. Furthermore, models like NoteLLM-2\cite{zhang2025notellm}, trained on user interaction data (clicks/likes), optimize for user interest rather than objective content relevance, making them ill-suited for strict semantic retrieval tasks where accuracy and consistency are paramount.

To address these limitations, we propose \textbf{UniNote}, a unified embedding model tailored for the complex demands of I2I retrieval as illustrated in Figure~\ref{fig:teaser}. UniNote bridges the gap between holistic encoding and fine-grained retrieval through a novel two-stage training paradigm. First, we introduce a \textbf{Contrastive Supervised Fine-tuning} stage. Unlike rigid dual-tower models, we develop a robust embedding model that compresses heterogeneous inputs into a unified vector space while supporting various I2I retrieval types. We introduce a suite of tailored retrieval settings to facilitate data-driven representation learning for complex multimodal content. This stage transforms an MLLM from a generative model into an embedding model by utilizing the last token as the final representation. Second, to overcome the ranking limitations of standard contrastive loss, we implement a \textbf{Relative Reranking via Reinforcement Learning} stage utilizing Group Relative Policy Optimization (GRPO)\cite{shao2024deepseekmath}. By constructing a hierarchical reward function that accounts for ranking positions and hard negatives, we directly optimize the embedding space for list-wise relevance. This enables the model to distinguish subtle content nuances without relying on computationally expensive generative rerankers, achieving high-precision ranking in a single pass. 

The main contributions are summarized as follows:
\begin{itemize}
    \item \textbf{Comprehensive I2I Task Suite:} We define ten distinct retrieval categories that span cross-modal alignment, information compression, and fine-grained local–global bidirectional retrieval. To support these, we construct a high-quality dataset with a robust hard-negative mining strategy.

    \begin{sloppypar}
    \item  \textbf{Unified Retrieval-Ranking Framework:} We propose \textbf{UniNote}, a unified model that bridges the gap between retrieval (embedding) and ranking (relevance). Through a novel two-stage training paradigm(Contrastive SFT followed by Relevance-aware Reinforcement Learning (via GRPO) ), we enable a single embedding to capture both coarse semantic spaces and fine-grained relevance.
    \end{sloppypar}

    \item \textbf{SOTA Performance \& Empirical Validation:} Extensive evaluations demonstrate that UniNote achieves state-of-the-art results across all ten I2I tasks. Ablation studies confirm that the two training stages synergize to balance embedding and ranking requirements in retrieval tasks.
    

    \item \textbf{Industrial Deployment Support \& Validation:} To meet large-scale industrial requirements, we adopt the MRL embedding scheme, allowing dimension selection based on cost–effectiveness trade-offs for flexible deployment. Production evaluations demonstrate that this approach maintains superior retrieval performance under practical constraints.
\end{itemize}

\section{Related Works}

\subsection{Multimodal Embedding}

Multimodal embedding aims to project heterogeneous data into a unified latent space. 
Early works, such as CLIP~\cite{radford2021learning}, BLIP~\cite{li2022blip}, and SigLIP
~\cite{zhai2023sigmoid}, pioneered the use of large-scale image-text pairs coupled with contrastive learning to achieve cross-modal alignment. These methods typically adopt a dual-tower architecture, where independent encoders project images and text into a shared manifold. While these models exhibit remarkable zero-shot generalization and have become the backbone for various downstream tasks\cite{deng2009imagenet,hendrycks2021many,hendrycks2021natural,zhou2017places,zhu2016visual7w,everingham2015pascal,mathew2021docvqa,lin13microsoft,liu2021visual,kazemzadeh2014referitgame}, their structural paradigm inherently limits them to single-modality inputs and often struggle with capturing fine-grained semantics or adhere to complex instructions due to their reliance on global, coarse-grained alignment.

The burgeoning demand for flexible and instruction-aware retrieval has catalyzed a shift toward leveraging MLLMs for representation learning. Most recent efforts have converged on a "last-token-as-embedding" paradigm. E5-V~\cite{jiang2024e5} employs prompt-guided MLLMs to compress multimodal context into a single descriptor, significantly improving performance in composite retrieval. UniME-V2~\cite{gu2025unime}  integrates multimodal contrastive learning with hard-negative mining via teacher-student distillation. Qwen3VL-Embedding~\cite{li2026qwen3} introduces a multi-stage training pipeline enriched with massive synthetic pairs and multi-task objectives, setting new benchmarks across diverse retrieval scenarios. By inheriting the reasoning capabilities of LLMs, these models transcend the limitations of traditional dual-towers, enabling a more nuanced understanding of complex multimodal queries. However, these models neglect the need to jointly capture global content representation and fine‑grained local retrieval in I2I tasks. In addition, they lack a unified retrieval-ranking framework that integrates  embedding with relevance ranking, thereby constraining their generality and consistent performance across diverse I2I retrieval scenarios.

\subsection{Embedding Model for Rerank}
For retrieval tasks, embedding models typically serve as the first-stage coarse retriever, while the second-stage fine-grained reranking is commonly handled by dedicated reranking models such as UniME-V2\cite{gu2025unime}, Lamra~\cite{liu2025lamra}, and Qwen3VL-Embedding\cite{li2026qwen3}.In UniME-V2's reranker, a generative MLLM is supervised through combined pairwise and listwise losses, and this paradigm is similarly adopted by Lamra and Qwen3VL-Embedding. 

Some approaches instead employ a unified embedding model for ranking; for instance, TaoSearchEmb\cite{liu2025taosearchemb} proposes Retrieval-GRPO, which follows the contrastive learning phase with reranking-oriented training. Specifically, after encoding the query and candidate sequences, ANN-based retrieval samples top-k items to serve as RL outcomes. However, its reward model operates as a black‑box with a single, rigid design, lacking generalizability and failing to ensure monotonic ranking consistency under conditional constraints defined by preferred recall quantity requirements, relative positional dependencies, and absolute positional dependencies in precise content‑matching retrieval tasks.



\section{Method}
\subsection{Task Definition}
\label{subsec:task_definiation}

\label{subsec:task_definition}
The I2I retrieval task is formally defined as follows: given a query $q \in \mathcal{Q}$ and a candidate set $\mathcal{C} = \{c_1, c_2, \dots, c_M\}$, the objective is to retrieve the candidate $c^*$ that is semantically relevant to $q$, where both $q$ and $c$ can represent items of any modality.

Within the Xiaohongshu platform, the predominant form of user‑generated content is referred to as a "Note", which serves as the primary medium for information sharing and interaction, constituting the core data format for our processing. A Note is primarily defined as:

\begin{equation}
    \mathcal{N} = \left( \{I_i\}_{i=1}^m,\{OCR_i\}_{i=1}^m, T_{title}, T_{body}\right)
\end{equation}

where $\{I_i\}_{i=1}^m$ denotes a set of $m$ images, and $OCR_i$ represents the text recognized within the $i$-th image $I_i$. $T_{title}$ and $T_{body}$ refer to the title and body text of the Note, respectively. In this work, videos are represented as a sequence of images without specialized identifiers.

The objective of UniNote is to establish a unified representation for each Note, replacing previous isolated content representations that typically learned independent embeddings for each modality. Specifically, UniNote aims to provide a unified embedding-based representation for multimodal data, facilitating both retrieval and relevance ranking within a single model. 




\subsection{Pipeline Overview}

\begin{figure*}
  \includegraphics[width=\textwidth]{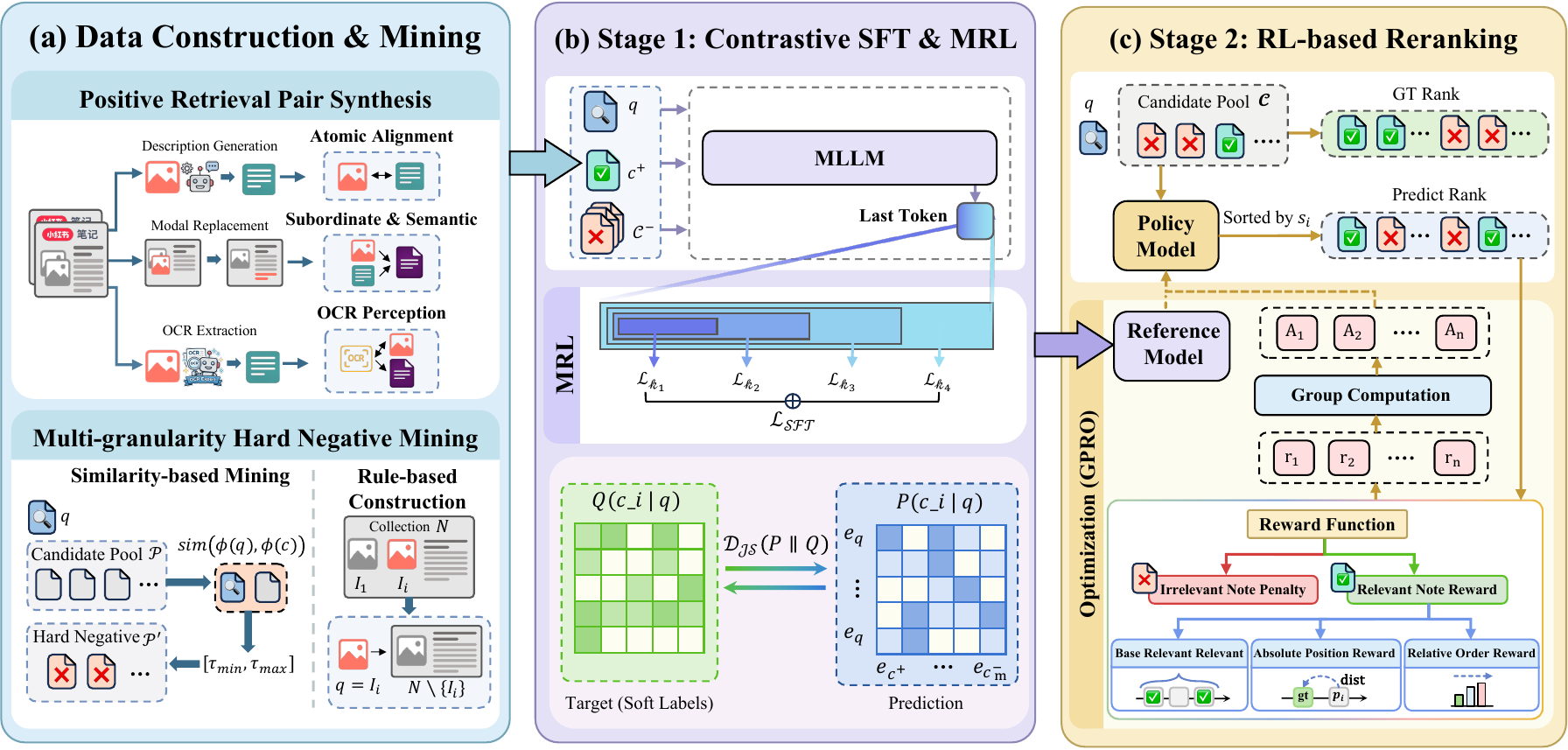}
  \caption{The training pipeline, where (a) presents the training data construction strategy designed for item-to-item retrieval, including retrieval sample pairs and hard negative mining strategy; (b) represents the training stage 1: contrastive supervised fine-tuning (SFT); and (c) represents the training stage 2: relative rerank via reinforcement learning (RL).}
  \label{fig:train_pipeline}
\end{figure*}

The training of UniNote follows a two-stage pipeline, comprising: (1) \textbf{Contrastive Supervised Fine‑Tuning}, and (2) \textbf{Relevance Reranking via  Reinforcement Learning}. As illustrated in Figure~\ref{fig:train_pipeline}. In the first stage (Sec.~\ref{sec:stage1}), we transform the generative MLLM into a high-quality embedding model. By leveraging the last-token hidden state as the representation. We adopt multi-granularity hard negative mining to improve the discrimination ability. In the second stage (Sec.~\ref{sec:stage2}), we bridge the gap between retrieval and reranking. By optimizing the model to perceive semantic overlap gradients through a reward-driven process, the embedding similarity can directly reflect the degree of content relevance, enabling a unified retrieval-ranking capability.

\subsection{Stage1: Contrastive Supervised Fine-tuning (Contrastive SFT)}
\label{sec:stage1}

This phase focuses on adapting the MLLM into a specialized embedding model by utilizing contrastive learning techniques. Through a data-driven methodology, the model develops the capacity to establish global representations while maintaining local-global mutual retrieval capabilities, which effectively compresses complex information into a dense latent representation. 

\subsubsection{\textbf{Training Data}} 
\label{subsubsec:training data}

In this stage, we need to construct matched training samples to support various retrieval sub-tasks within Note scenarios, while simultaneously enhancing the discriminative power of the model through hard negative mining. 

In real-world application contexts, the requirements for Note retrieval scenarios are categorized into five meta-tasks, encompassing a total of ten specific retrieval tasks as shown in Table~\ref{tab:dataset_Taxonomy}. (1) \textbf{Atomic Alignment} represents a fundamental capability, establishing the mapping between a single image and its corresponding textual description to enable basic cross-modal search. (2) \textbf{Subordinate Retrieval} refers to application scenarios in which local content, such as specific images or text segments, is utilized to retrieve the complete Note. (3) \textbf{Semantic Extraction} functions as a complement to subordinate retrieval and is critically important in applications such as Note risk assessment. (4) \textbf{OCR Perception} addresses the prevalence of text-rich images on Xiaohongshu, creating the need to directly retrieve relevant images and Notes through OCR-based capabilities. (5) \textbf{Content Relevance} measures the degree of content relevance between different Notes.

\begin{table}[!h]
    \centering
    \caption{I2I Retrieval Tasks Taxonomy.}
    \label{tab:dataset_Taxonomy}
    \small
    \begin{tabular}{@{}ll@{}}
    \toprule
    \textbf{Meta Task} & \textbf{Search Type} \\ 
    \midrule
    Atomic Alignment & image $\leftrightarrow$ text \\
    Subordinate Retrieval & image / text $\to$ note \\
    Semantic Extraction & note $\to$ image / text \\
    OCR Perception & image $\leftrightarrow$ ocr, ocr $\to$ note \\
    Content Relevance & note $\to$ note \\
    \bottomrule
    \end{tabular}
\end{table}


The positive retrieval pair construction begins with high-quality notes collected from Xiaohongshu. First, based on each image \( I_i \) in note \( \mathcal{N} \), we use an annotator model \( \phi_{\text{ann}} \) to generate semantically aligned image-text pairs \( (I_i, S_{\text{desc}}^i) \) for Atomic-level Alignment. Then, for \texttt{Subordinate Retrieval} and \texttt{Semantic Extraction}, we leverage the inherent subordinate relationship between notes and their constituent elements, with ground-truth pairs defined as $(I_i,\mathcal{N})$ where $I_i \in \mathcal{N}$. In order to prevent the model from exploiting "low-level shortcuts" where it might match pixels or local features instead of understanding semantics, we introduce a \textbf{Modal Replacement} mechanism that substitutes the original image with its textual description:
\begin{equation}
    (I_i, \mathcal{N}'), \quad \text{where} \quad \mathcal{N}' = \{S_{\text{desc}}^i \} \cup \mathcal{N} \setminus \{I_i\} 
\end{equation}
This formulation also supports text-query variants in the form of $(S_{\text{desc}}^i, \mathcal{N})$. For \texttt{OCR Perception.} We employ expert OCR models to extract textual content embedded within images, yielding pairs $(I_i, S_{\text{ocr}}^i)$. These pairs provide positive samples for the I2OCR, OCR2I, and OCR2Note tasks.

\paragraph{Multi-granularity Hard Negative Mining.} The embedding model aims to pull related samples closer and push unrelated ones apart via contrastive learning. In this context, hard negative samples are critical for enhancing the discriminative capability of the model. To facilitate robust hard negative mining across diverse scenarios, we utilize both relevance score techniques \cite{gu2025unime,zheng2023judging,chen2024mllm} and counterfactual hard negatives.

Given a query $q$ and its corresponding positive samples $c^{+}$, we aim to construct a set of hard negative samples $\mathcal{C}^{-} = \{c_{1}^{-},\dots ,c_{k}^{-}\}$. Following the paradigm of UniME-V2, our approach integrates global mining with soft supervision. Let $\mathcal{P}$ denote the global candidate pool and $\phi(\cdot)$ represent a reference encoder. The subset of valid hard negatives $\mathcal{P}' \subset \mathcal{P}$ is defined as:
\begin{equation}
\mathcal{P}' = \{ c \in \mathcal{P} \mid \tau_{\min} \leq \operatorname{sim}(\phi(q), \phi(c)) \leq \tau_{\max} \},
\end{equation}
where $[\tau_{\min},\tau_{\max}]$ defines the similarity range that identifies informative negatives while excluding false negatives. The similarity score $s = \operatorname{sim}(\phi(q),\phi(c^{-}))$ for $c^{-}\in\mathcal{P}'$ is subsequently utilized as a soft supervision signal.

To address scenarios where candidates are composite structures (e.g., \texttt{I2Note} or \texttt{T2Note}) rather than monolithic entities, we refine the scoring mechanism. Since standard encoders often fail to simultaneously capture global semantics and local fine-grained details, we define a refined soft score $s'$ using a MaxSim operation:
\begin{equation}
s'(q, c^-) = \max_{e \in c^-} \operatorname{sim}(q, e),
\end{equation}
where $e$ denotes the constituent elements of the composite candidate $c^{-}$. This refined soft score $s'$
subsequently serves as the supervision target for distribution alignment.

Furthermore, we incorporate heuristic rules based on the membership relation $(I_{i},\mathcal{N})$, where $I_{i}\in \mathcal{N}$ . We construct hard negative pairs via the following set subtraction:
\begin{equation}
(q, c^-_{\text{rule}} ) = \bigl(I_{i},\, \mathcal{N} \setminus \{I_{i}\}\bigr).
\end{equation}
This formulation ensures that the negative sample maintains maximum content overlap with the positive candidate context while strictly precluding any valid membership relationship.

\subsubsection{\textbf{Optimization Goal}}
We minimize the divergence between the embedding-based distribution $P$ and the MLLM-derived soft-label distribution $Q$. Given a candidate set $\Omega_c = \{c^+, c_1^-, \dots, c_m^-\}$, the distributions are formulated as:
\begin{equation}
    Q(c_i | q) = \frac{\exp(s_{q, c_i} / \tau)}{\sum_{c_j \in \Omega_c} \exp(s_{q, c_j} / \tau)}
\end{equation}
\begin{equation}
    P(c_i | q) = \frac{\exp(\cos(e_q, e_{c_i}) / \tau)}{\sum_{c_j \in \Omega_c} \exp(\cos(e_q, e_{c_j}) / \tau)}
\end{equation}
To ensure learning robustness and symmetry, we employ the Jensen-Shannon (JS) divergence as the objective:
\begin{equation}
    \mathcal{D}_{JS}(P \parallel Q) = \frac{1}{2} \text{KL}(P \parallel Q) + \frac{1}{2} \text{KL}(Q \parallel P)
\end{equation}

\subsection{Stage 2: Relative Reranking via Reinforcement Learning}
\label{sec:stage2}

In this stage, we utilize Note2Note retrieval training data to enhance the perception of content relevance and the ranking capability of the model. We define the relevance between two notes as the degree of content overlap in our training settings. To construct training samples, we consider a given note $\mathcal{N}$ and partition both its visual and textual components into two equal sub-notes, $\mathcal{N}_A$ and $\mathcal{N}_B$. This partition satisfies the conditions $\mathcal{N}_A \cap \mathcal{N}_B = \emptyset$ and $\mathcal{N}_A \cup \mathcal{N}_B = \mathcal{N}$, ensuring that $\mathcal{N}_A$ and $\mathcal{N}_B$ initially have no content overlap.

With $\mathcal{N}_A$ serving as the query $q$, we construct a candidate sequence with incrementally decreasing content overlap as follows: $L_{rel} = [ \mathcal{N}_B \cup \{I^A_1, I^A_2, \dots\},\dots, \mathcal{N}_B \cup \{I^A_1\}, \mathcal{N}_B]$. Furthermore, we introduce a noise sequence, denoted as $\{\mathcal{N}_{noise},\dots \}$. The final ranking list $L_{rank}$ is then systematically constructed by integrating the content‐relevant sequence with the noise sequence, defined as: $L_{rank} = [\mathcal{N}_B \cup \{I^A_1, I^A_2, \dots,I^A_M\},\dots, \mathcal{N}_B \cup \{I^A_1\}, \mathcal{N}_B, \mathcal{N}^{noise}_1,\mathcal{N}^{noise}_2,..]$. Here, M is the number of content‐relevant notes in $L_{rank}$  for the given query.


In contrast to existing frameworks such as UniME-V2 and Qwen3VL‐Embedding, which typically utilize a separate model for the reranking stage, we unify representation learning and reranking within a single model architecture. We use GRPO framework for training, as the reranking process inherently aligns with the concept of group-based rewards. This process is illustrated in Figure~\ref{fig:train_pipeline}(c), consisting of three stages: ranking prediction, reward design, and optimization.

\subsubsection{\textbf{Candidates Selection}} 
This step simulates the group sampling process within the GRPO framework. Specifically, the top-$G$ (G>M) notes retrieved by the embedding model are treated as the candidate group. By computing the similarity scores, we establish a descending ranking order: $P = [p_1, p_2, \dots, p_G]$. Meanwhile, the corresponding ground truth ranking is denoted as $V = [g_1, g_2, \dots, g_G]$.


\subsubsection{\textbf{Reward Design}} The reward function (Figure  ~\ref{fig:rewardfunc}) is designed to strength two specific capabilities of the model: the ability to discriminate between relevant and irrelevant notes, and the ability to rank notes according to their relevance. Consequently, the reward function $R$ is constructed across four dimensions: irrelevance penalty, relevance reward, absolute position reward, and relative position reward. We design a multi-dimensional reward function to supervise the quality of ranking across three levels of granularity:

\begin{figure}
    \centering
  
\includegraphics[width=1.0\linewidth]{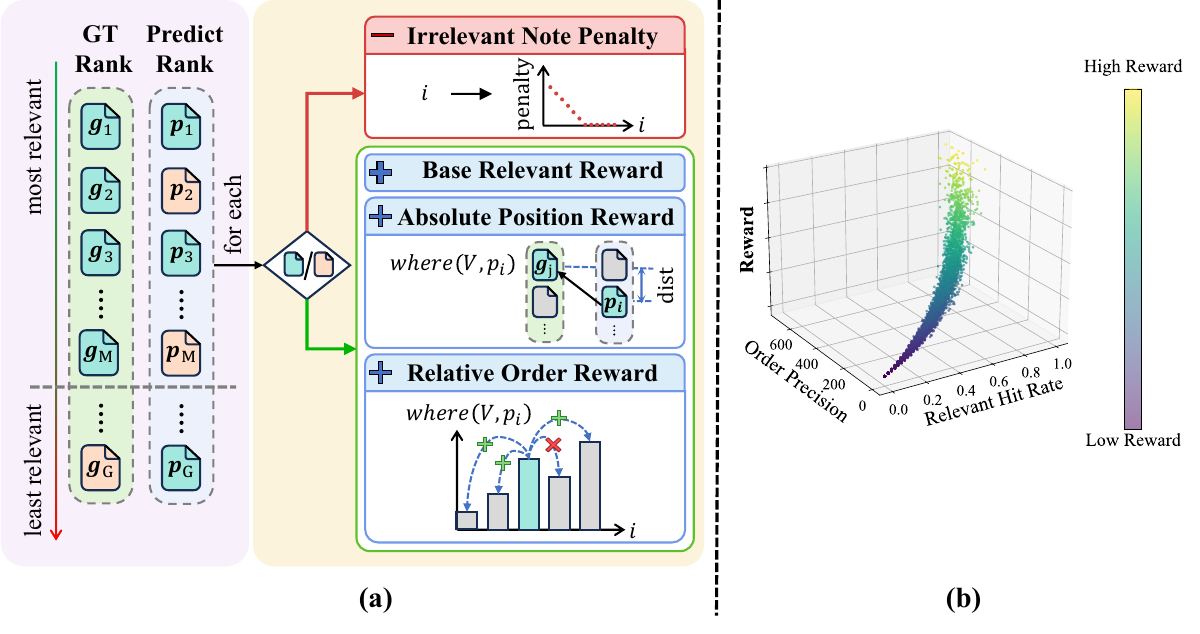}
    \caption{Schematic illustration of the reward function. (a) The reward function comprises four components: Irrelevant Note Penalty, Base Relevant Reward, Absolute Position Reward, and Relative Order Reward. (b) The reward distribution for simulated prediction ranking is depicted. The reward increases in conjunction with the relevant hit rate and order precision, where the relevant hit rate plays a more decisive role in the overall reward value.
 }
   \label{fig:rewardfunc}
\end{figure}

\begin{itemize}
\item Irrelevant Note Penalty: We penalize noise notes $p_i \notin L_{rel}$ that are ranked above any relevant note ($\exists p_j \in L_{rel}, j > i$). The penalty escalates as $i$ approaches 1, reflecting the increased harm of ranking irrelevant content highly.

\item Base Relevant Reward : 
In our business context, we prioritize maximizing the recall of relevant samples in $P$ over the ranking order of relevant notes. Therefore, we assign a fixed base reward $R_{\mathrm{base}}$ to any note where $p_{i} \in L_{\mathrm{rel}}$.

\item Absolute Position Reward: To minimize the displacement between predicted and ground-truth positions, we define a reward based on the distance $\mathrm{dist}=|p_{i}-where(V,p_i)|$. A smaller distance implies a more accurate absolute position.
\item Relative Order Reward: Since absolute distance alone cannot capture the global consistency of the sequence, we calculate a relative order reward inspired by the concept of concordant pairs. For each relevant note $p_{i}$, we count the number of other relevant notes $p_{j}$ that are correctly ordered relative to $i$.
\end{itemize}

\noindent The reward $r_{i}$ for the note at position $i$ is formulated as follows:
\begin{equation}
    R_{\mathrm{abs}}(i,p_i)=1-\frac{|p_{i}-where(V,p_i)|}{G}
\end{equation}

\begin{equation}
    R_{\mathrm{rel}}(i, P) = mean(
\mathbb{I}\bigl[ (p_i-p_j)(where(V,p_i)-where(V,p_j) ) > 0  ]\bigr)
\end{equation}

\begin{equation}
r_i =
\begin{cases}
C_{\mathrm{pen}} \cdot \left(1 + \dfrac{G-i}{G}\right) & \text{if } p_i \notin L_{rel} \text{  and } \exists p_j \in L_{rel} ,j >i \\[8pt]
0  & \text{if } p_i \notin L_{rel} \text{  and } \nexists p_j \in L_{rel} ,j >i\\[8pt]
R_{\mathrm{base}} + R_{\mathrm{abs}}(p_i, i) + R_{\mathrm{rel}}(p_i, P) & \text{if } p_i \in L_{rel}
\end{cases}
\end{equation}
where $C_{\mathrm{pen}}$ and $R_{\mathrm{base}}$ are fixed hyperparameters,

\subsubsection{\textbf{Optimization}} We compute the intra-group advantage scores for the predicted candidates. The advantage $A_{i}$ for each candidate in the group is computed as:
\begin{equation}
A_i = \frac{r_i - mean(r) }
              {std(r) + \epsilon}
\end{equation}

Following the GRPO framework, the final loss function $\mathcal{L}_{GRPO}$ is formulated as:
\begin{equation}
\mathcal{L}_{GRPO} = \mathbb{E} \left[ \frac{1}{G} \sum_{i=1}^{G} \left( A_i \cdot \log \pi_{\theta} ( sim(q, p_i) \mid q, p_i ) - \beta \mathbb{D}_{KL} ( \pi_{\theta} \parallel \pi_{ref} ) \right) \right]
\end{equation}
where $\pi_{\theta}$ denotes the policy model, which corresponds to our embedding model.

\subsection{MRL Support}

To support diverse deployment constraints, we integrate \textbf{MRL} into our alignment framework. We define a set of nested dimensions $\mathcal{K} = \{64, 512, 1024, 4096\}$. For each dimension $k \in \mathcal{K}$, we minimize the divergence between the embedding-based distribution $P^{(k)}$ and the MLLM-derived soft-label distribution $Q$. The final total loss $\mathcal{L}_{SFT}$ is the  sum across all nested dimensions:
\begin{equation}
    \mathcal{L}_{SFT} = \sum_{k \in \mathcal{K}}   \mathcal{D}_{JS}\left(P^{(k)} \parallel Q\right)
\end{equation}
This ensures that lower-dimensional truncations maintain high retrieval accuracy while inheriting the semantic depth of the full-dimensional representation.

\section{Experiment Results}
\subsection{Evaluation Settings}

We utilize Qwen3VL-8B-Instruct as the foundation framework for UniNote. During the RL stage, we employ the GRPO with the Adam for optimization. To address the lack of text in raw training data, we leverage Qwen3VL-8B-Instruct to generate descriptive captions and Qwen3VL-8B-Embedding for hard negative mining. The training is conducted on 8\ H800 GPUs.


The evaluation is performed on a comprehensive multimodal dataset consisting of 66k notes and over 500k items, covering image, text, and OCR modalities. We evaluate models across the five meta-tasks defined in Table ~\ref{tab:dataset_Taxonomy}. To assess performance, we primarily utilize Recall-based metrics. Specifically, we denote the metric as $\mathrm{R}^*@\mathrm{K}$ for the Note2Note task, which involves multiple ground-truth candidates per query, and as $\mathrm{R}@\mathrm{K}$ for the remaining four categories, where each query corresponds to a unique positive sample. Detailed formulations and evaluation protocols are provided in the Appendix.

\begin{table*}[!ht]
    \centering
    \caption{Performance comparison of different embedding models (\%).}
    \label{tab:main_result}
    \resizebox{\textwidth}{!}{
    \begin{tabular}{llcc ccc ccc >{\columncolor{gray!10}}c >{\columncolor{gray!10}}c >{\columncolor{gray!10}}c}
    \toprule
    \multirow{2}{*}{\textbf{Category}} & \multirow{2}{*}{\textbf{Search Type}} & \multirow{2}{*}{\textbf{n\_queries}} & \multirow{2}{*}{\textbf{n\_targets}} & \multicolumn{3}{c}{\textbf{RzenEmbed}} & \multicolumn{3}{c}{\textbf{Qwen3VL-Embed.}} & \multicolumn{3}{c}{\cellcolor{gray!10}\textbf{UniNote (Ours)}} \\
    \cmidrule(lr){5-7} \cmidrule(lr){8-10} \cmidrule(lr){11-13}
    & & & & R@1 & R@5 & R@10 & R@1 & R@5 & R@10 & R@1 & R@5 & R@10 \\ 
    \midrule
    
    \multirow{2}{*}{\makecell[l]{Atomic \\ Alignment}}
    & I2T & 50w & 48w & 63.6 & 82.7 & 87.1 & 56.0 & 76.1 & 81.6 & \textbf{75.2} & \textbf{90.2} & \textbf{92.9} \\ 
    & T2I & 48w & 50w & 69.5 & 86.6 & 90.2 & 52.9 & 72.6 & 78.4 & \textbf{74.0} & \textbf{89.1} & \textbf{92.0} \\ 
    \midrule
    
    \multirow{2}{*}{\makecell[l]{Subordinate\\Retrieval}} 
    & I2Note & 50w & 13w & 37.9 & 59.6 & 65.5 & 41.2 & 61.1 & 67.4 & \textbf{93.7} & \textbf{100} & \textbf{100} \\ 
    & T2Note & 48w & 13w & 32.2 & 53.8 & 59.8 & 39.6 & 62.4 & 68.5 & \textbf{53.1} & \textbf{72.0} & \textbf{77.3} \\ 
    \midrule
    
    \multirow{2}{*}{\makecell[l]{Semantic\\Extraction}} 
    & Note2I & 6.6w & 6.6w & 51.2 & 68.3 & 73.6 & 64.3 & 80.8 & 85.3 & \textbf{90.2} & \textbf{91.6} & \textbf{92.7} \\ 
    & Note2T & 6.6w & 6.6w & 43.1 & 61.5 & 67.4 & 44.6 & 63.3 & 69.3 & \textbf{64.2} & \textbf{83.8} & \textbf{87.7} \\ 
    \midrule
    
    \multirow{3}{*}{\makecell[l]{OCR\\Perception}} 
    & OCR2Note & 18w & 13w & 33.2 & 55.2 & 60.6 & 37.7 & 60.6 & 66.4 & \textbf{62.1} & \textbf{80.5} & \textbf{84.5} \\ 
    & I2OCR & 20w & 18w & 70.4 & 84.0 & 86.9 & \textbf{79.0} & \textbf{91.3} & \textbf{93.3} & 53.5 & 66.7 & 69.9 \\ 
    & OCR2I & 18w & 20w & 55.8 & 71.2 & 75.5 & 64.3 & 79.0 & 82.5 & \textbf{82.8} & \textbf{90.9} & \textbf{92.5} \\ 
    \midrule



    \multirow{2}{*}{\makecell[l]{Content\\Relevance}} 
    & \multirow{2}{*}{Note2Note} & \multirow{2}{*}{3.7w} & \multirow{2}{*}{37w} 
    & $\mathrm{R}^*@1$ & $\mathrm{R}^*@5$ & $\mathrm{R}^*@10$ & $\mathrm{R}^*@1$ & $\mathrm{R}^*@5$ & $\mathrm{R}^*@10$ & $\mathrm{R}^*@1$ & $\mathrm{R}^*@5$ & $\mathrm{R}^*@10$ \\ 
    
    \hhline{~~~~|---||---||---||}
    
    & & & & 15.8 & 79.3 & 98.6 & \textbf{15.9} & \textbf{79.6} & 99.4 & \textbf{15.9} & 79.3 & \textbf{99.5} \\
    \bottomrule
    
    \end{tabular}
    }
\end{table*}

\subsection{Comparison with State-of-the-Arts}

 We selected RzenEmbedding \cite{jian2025rzenembed} and Qwen3VL-Embedding-8B as our primary baselines, as both have achieved state-of-the-art (SOTA) performance in general retrieval tasks. As illustrated in Table~\ref{tab:main_result}, while existing SOTA methods exhibit robust performance in fundamental cross-modal alignment tasks (e.g., atomic alignment), but they falter in Subordinate Retrieval and Semantic Extraction. We attribute this to their training paradigms, which rely heavily on single image-text pairs. Consequently, these models lack the capacity to model the intricate global-local relationships inherent in complex, multimodal compositions, leading to an inability to balance coarse-grained global representation with fine-grained local retrieval.

Furthermore, in the OCR perception task, both RzenEmbed and Qwen3VL-Embedding-8B demonstrate robust I2OCR retrieval capabilities. However, the inverse capability of using OCR to retrieve corresponding images is significantly weaker. This performance gap is further amplified when attempting to retrieve entire notes based on OCR queries. In contrast, UniNote employs a data-driven approach to strike an effective balance across these diverse tasks. UniNote achieves the best performance across all tasks with the exception of I2OCR. It not only facilitates effective cross-modal alignment but also yields substantial improvements over RzenEmbed and Qwen3VL-Embedding-8B in local-to-global bidirectional retrieval tasks, such as I2Note, Note2I, and OCR2Note.

For the Note2Note retrieval task, we constructed a candidate pool by mixing irrelevant "distractor" notes with M relevant notes, requiring the model to identify and rank the top-M candidates. We found that both Rzen and Qwen3VL-Embedding performed respectably in this setting. This may be due to the nature of the task, where local overlap serves as a salient feature that is relatively easy for encoders to capture.

\subsection{Ablation Result} 
To evaluate the efficacy of our proposed two-stage training framework, we conducted a series of ablation studies across four distinct configurations, as summarized in Table~\ref{tab:ablation study}.

We evaluate the efficacy of our hard negative mining strategy in SFT stage: First, we examine the rand SFT variant, where negative samples are randomly selected from the global pool. This approach tends to introduce false negatives and suffers from an excessive semantic gap between positive and negative pairs. Such easy negatives fail to provide informative optimization gradients, thereby preventing the model from learning highly discriminative representations. Consequently, the performance of rand SFT on four meta-tasks is significantly lower than that of UniNote. This result demonstrates that purely random sampling is insufficient for capturing fine-grained cross-modal nuances.

Next, we exclude hard negatives generated through rule-based heuristics. The substantial performance improvement of without rule-based mining compared to rand mining validates the significant value of global hard negative mining and soft supervision. Building upon this, the further performance gains achieved by incorporating counterfactual hard negatives demonstrate that this strategy forces the model to distinguish more subtle semantic differences, thereby increasing the confidence of the model in identifying subordinate relationships.

Finally, we evaluate the impact of the RL stage on the Note2Note task. As illustrated in Table~\ref{ablation on RL}, the introduction of the RL stage enhances both the precision and recall of the model for the Note2Note task. Specifically, the R@5 and P@5 metrics increase by 1.4\% and 3.4\%, respectively. The results indicate that, across various retrieval and matching tasks, UniNote achieves a higher relevant-sample recall rate than the SFT mode when the number of retrieved samples is controlled to be identical.

\begin{table*}[!ht]
    \centering
    \setlength{\tabcolsep}{5pt} 
    \caption{Hard Negative Sample Ablation Results (\%).} 
    \label{tab:ablation study}
    \begin{tabular}{llccccc}
    \toprule
    \textbf{Method} & \textbf{Metric} & \makecell{Atomic Alignment} & \makecell{Subordinate Retrieval} & \makecell{Semantic Extraction} & \makecell{OCR Perception} &\textbf{Total} \\ 
    \midrule
    \multirow{3}{*}{rand} 
        & R@1  & 57.6 & 66.2 & 62.6 & 42.8 & 57.3 \\ 
        & R@5  & 76.5 & 79.5 & 81.6 & 56.4 & 73.5 \\ 
        & R@10 & 81.7 & 83.1 & 86.0 & 60.2 & 77.7 \\ 
    \midrule 
    \multirow{3}{*}{\shortstack[l]{UniNote\\(w/o Rule)}} 
        & R@1  & 71.9 & 74.1 & 73.9 & 47.0 & 66.7 \\ 
        & R@5  & 87.6 & 84.4 & 88.6 & 57.3 & 79.5 \\ 
        & R@10 & 90.8 & 87.2 & 91.3 & 59.4 & 82.2 \\ 
    \midrule
    \multirow{3}{*}{\shortstack[l]{UniNote}} 
        & R@1  & \textbf{74.6 (\hgreen{+17.0})} & \textbf{76.4 (\hgreen{+10.2})} & \textbf{77.2 (\hgreen{+14.6})} & \textbf{66.1 (\hgreen{+23.3})} & \textbf{73.6 (\hgreen{+16.3})} \\ 
        & R@5  & \textbf{89.6 (\hgreen{+13.1})} & \textbf{86.0 (\hgreen{+6.5})} & \textbf{87.7 (\hgreen{+6.1})} & \textbf{79.3  (\hgreen{+22.9})} & \textbf{85.7 (\hgreen{+12.2})} \\ 
        & R@10 & \textbf{92.5 (\hgreen{+10.8})} & \textbf{88.6 (\hgreen{+5.5})} & \textbf{90.2 (\hgreen{+4.2})} & \textbf{82.3 (\hgreen{+22.1})} & \textbf{88.4 (\hgreen{+10.7})} \\ 
    \bottomrule
    \end{tabular}
\end{table*}

\begin{table}[!ht]
    \centering
    \setlength{\tabcolsep}{2pt} 
    \caption{Ablation study on RL module (\%).} 
    \label{ablation on RL}
    
    \begin{tabular}{l ccc | ccc}
    \toprule
        Model & R@1 & R@5 & R@10 & P@1 & P@5 & P@10  \\ 
    \midrule
        \makecell{UniNote \\ (w/o RL) }
         & 15.7 & 77.9 & 97.9 & 91.7  & 92.6 & 61.8  \\ 
    \midrule
        UniNote &
        \textbf{\makecell{15.9 \\ (\hgreen{+0.2})}} & 
        \textbf{\makecell{79.3 \\ (\hgreen{+1.4})}} & 
        \textbf{\makecell{99.5 \\ (\hgreen{+1.6})}} & 
        \textbf{\makecell{96.0 \\ (\hgreen{+4.3})}} & 
        \textbf{\makecell{96.0 \\ (\hgreen{+3.4})}} & 
        \textbf{\makecell{64.5 \\ (\hgreen{+2.7})}} \\ 
    \bottomrule
    \end{tabular}
\end{table}





\subsection{MRL Embedding Performance } \label{subsec:matryoshka}

\begin{figure}
    \centering
    \includegraphics[width=1.0\linewidth]{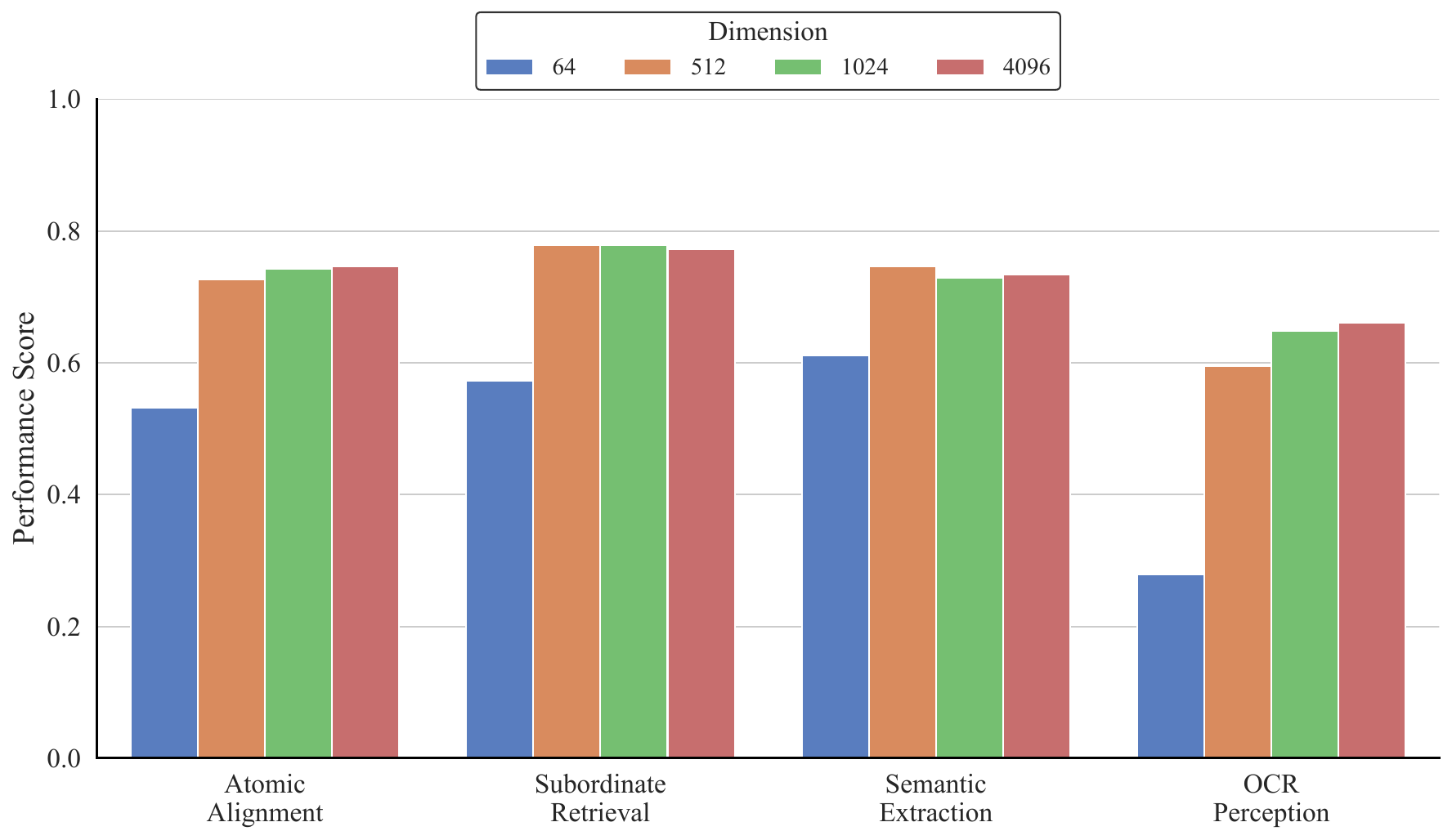}
    \caption{Impact of feature dimensions on efficiency and recall with Matryoshka embeddings.}
    \label{fig:matryoshka_analysis}
\end{figure}

To balance retrieval performance and storage overhead, we incorporate Matryoshka Loss to support multi-dimensional outputs. Figure~\ref{fig:matryoshka_analysis} illustrates the impact of feature dimensions (64, 512, 1024, and 4096) on efficiency and Recall. The results demonstrate that the overall performance exhibits an upward trend as the feature dimension increases. Specifically, at a dimension of 512, the performance of most retrieval tasks approximates that of the full-dimensional features. Although a decline in performance is observed at a dimension of 64, the model retains 70\% of the capability in tasks such as Atomic Alignment, Subordinate Retrieval, and Semantic Extraction. Consequently, the Matryoshka Embedding enables flexible deployment, catering to the requirements of diverse application scenarios.

\section{Online Deployment}

\begin{figure}
    \centering
    \includegraphics[width=1.0\linewidth]{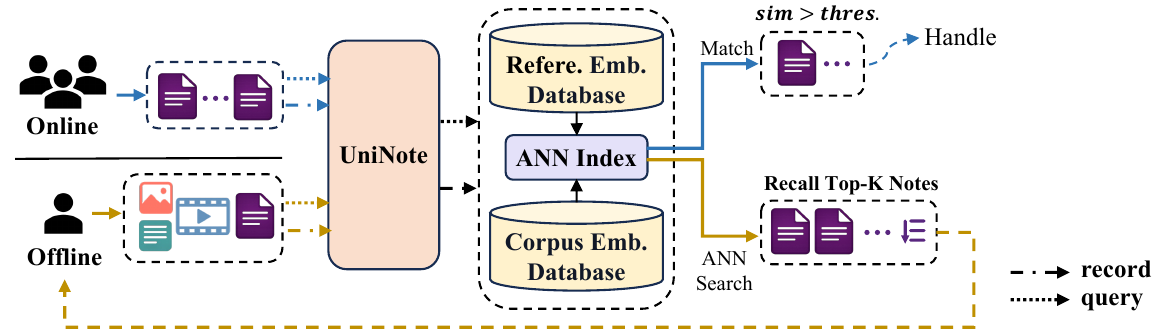}
    \caption{Online deployment pipeline. The online mode refers to the process in which massive user-generated notes are stored into the corpus and simultaneously used as queries to perform matching verification against the reference small-sample library. The offline mode refers to the process where internal operations personnel update the reference small-sample library with risk-related data and perform retrieval and recall from the historical corpus}
    \label{fig:online_test}
\end{figure}

We conducted deployment testing of UniNote following the setup illustrated in Figure~\ref{fig:online_test}, operating in two modes: online and offline, corresponding to real-time production traffic and manually triggered tasks, respectively.

In the online mode, high-concurrency traffic streams are processed in real time by UniNote for embedding extraction and subsequent storage. Simultaneously, the Approximate Nearest Neighbor (ANN) index is updated to incorporate historical data. In parallel, the incoming traffic is matched against a pre-built small-scale sample repository for validation, and any matched notes are subsequently processed according to the downstream task requirements.

In the offline mode, operations personnel can initiate two categories of tasks. The first involves processing multimodal and multi-type samples (e.g., image, text, video, and note) for embedding extraction via UniNote, followed by storage and index updates—corresponding to the construction and maintenance of the aforementioned sample repository in the online mode. The second category enables retrieval from the historical data index using any given item as a query, returning the Top‑K most relevant samples. Building upon the above definitions, we classify the evaluation tasks into two categories: (1) high‑request, small‑gallery verification, and (2) low‑request, large‑gallery retrieval.

\paragraph{Online mode (high‑request, small‑gallery verification).} 
We evaluate UniNote in an online high-concurrency setting, matching daily incoming notes against a safety policy gallery. Over 7 consecutive days, 10\% of daily traffic was randomly sampled for A/B testing. The baseline is a production CLIP-based image–text model trained on domain-specific data. The gallery contains three risk categories—images, videos, and notes—with 50 samples each (150 total). Similarity thresholds were tuned per category to match the baseline’s retrieval count, and professional annotators verified results. UniNote achieved 85.6\%, 91.2\%, and 93.6\% recall retention in Note2Image, Note2Video, and Note2Text tasks using a single embedding extraction, while the baseline Image-to-Image method required 9.2× more storage and compute. UniNote thus delivers high recall with substantially lower resource cost.


\paragraph{Offline mode (low‑request, large‑gallery retrieval).} 
In the back-checking scenario, known interest images (or notes and other related items) are used as queries to retrieve a collection of notes that contain content relevant to the given interest, from a large-scale historical note repository. In this experimental setting, we likewise employ the existing online CLIP-based image–text alignment model as the baseline and select one week of online data as the historical note collection. For the query configuration, we adopt the ANN search method and fix the number of returned results to the top‑10 notes per query. A set of 1K images of various types was used as queries, collectively yielding 10K note retrieval results for annotation. After deduplication of identical notes, UniNote achieved a 23.5\% gain in relevant sample recall. This demonstrates that retrieval methods based on note sub-content tend to result in more duplicate recalls under identical computational cost, further highlighting the necessity of unified representation and retrieval.

The above two online experiments demonstrate the efficiency and practical viability of UniNote in industrial retrieval tasks.


\section{Conclusion and Future Work}
In this work, we address the item-to-item (I2I) retrieval challenge in multimodal scenarios, aiming to balance global representation and fine-grained retrieval while unifying multimodal representation and relevance search in a shared embedding space. Using real Xiaohongshu note data, we designed ten retrieval tasks of varying granularity and transformed them into high-quality embeddings via a Multimodal Large Language Model (MLLM). Through cross-modal alignment combined with similarity scoring and counterfactual hard negative mining, UniNote achieved substantial gains in I2I retrieval. In the Note2Note task, we introduced GRPO-based reinforcement learning, proposing a robust relevance-ranking reward that advances the unification of representation and ranking, with scalability to tasks such as image-to-note (I2Note). This approach offers an efficient, practical solution for industrial-scale heterogeneous item retrieval, and paves the way for leveraging reinforcement learning to further boost retrieval and ranking across diverse scenarios.

\bibliographystyle{ACM-Reference-Format}
\balance 
\bibliography{sample-base}

@String{Computing = "Computing" }

@String{Computer = "{IEEE} Computer" }

@String{Springer = "Springer-Verlag" }

@article{liu2024multimodal,
  title={Multimodal recommender systems: A survey},
  author={Liu, Qidong and Hu, Jiaxi and Xiao, Yutian and Zhao, Xiangyu and Gao, Jingtong and Wang, Wanyu and Li, Qing and Tang, Jiliang},
  journal={ACM Computing Surveys},
  volume={57},
  number={2},
  pages={1--17},
  year={2024},
  publisher={ACM New York, NY}
}

@inproceedings{wei2019mmgcn,
  title={MMGCN: Multi-modal graph convolution network for personalized recommendation of micro-video},
  author={Wei, Yinwei and Wang, Xiang and Nie, Liqiang and He, Xiangnan and Hong, Richang and Chua, Tat-Seng},
  booktitle={Proceedings of the 27th ACM international conference on multimedia},
  pages={1437--1445},
  year={2019}
}

@inproceedings{wu2022mm,
  title={Mm-rec: Visiolinguistic model empowered multimodal news recommendation},
  author={Wu, Chuhan and Wu, Fangzhao and Qi, Tao and Zhang, Chao and Huang, Yongfeng and Xu, Tong},
  booktitle={Proceedings of the 45th international ACM SIGIR conference on research and development in information retrieval},
  pages={2560--2564},
  year={2022}
}

@article{xu2024large,
  title={Large language models for generative information extraction: A survey},
  author={Xu, Derong and Chen, Wei and Peng, Wenjun and Zhang, Chao and Xu, Tong and Zhao, Xiangyu and Wu, Xian and Zheng, Yefeng and Wang, Yang and Chen, Enhong},
  journal={Frontiers of Computer Science},
  volume={18},
  number={6},
  pages={186357},
  year={2024},
  publisher={Springer}
}

@inproceedings{yuan2024rethinking,
  title={Rethinking multimodal content moderation from an asymmetric angle with mixed-modality},
  author={Yuan, Jialin and Yu, Ye and Mittal, Gaurav and Hall, Matthew and Sajeev, Sandra and Chen, Mei},
  booktitle={Proceedings of the IEEE/CVF winter conference on applications of computer vision},
  pages={8532--8542},
  year={2024}
}

@article{li2025towards,
  title={Towards Trustworthy Multimodal Moderation via Policy-Aligned Reasoning and Hierarchical Labeling},
  author={Li, Anqi and Jin, Wenwei and Tong, Jintao and Qin, Pengda and Li, Weijia and Lu, Guo},
  journal={arXiv preprint arXiv:2508.03296},
  year={2025}
}

@article{arya2024mscmgtb,
  title={MSCMGTB: A Novel Approach for Multimodal Social Media Content Moderation Using Hybrid Graph Theory and Bio-Inspired Optimization},
  author={Arya, Premnarayan and Pandey, Amit Kumar and Patro, S Gopal Krishna and Tiwari, Kretika and Panigrahi, Niranjan and Naveed, Quadri Noorulhasan and Lasisi, Ayodele and Khan, Wahaj Ahmad},
  journal={IEEE Access},
  volume={12},
  pages={73700--73718},
  year={2024},
  publisher={IEEE}
}

@inproceedings{radford2021learning,
  title={Learning transferable visual models from natural language supervision},
  author={Radford, Alec and Kim, Jong Wook and Hallacy, Chris and Ramesh, Aditya and Goh, Gabriel and Agarwal, Sandhini and Sastry, Girish and Askell, Amanda and Mishkin, Pamela and Clark, Jack and others},
  booktitle={International conference on machine learning},
  pages={8748--8763},
  year={2021},
  organization={PmLR}
}

@inproceedings{li2022blip,
  title={Blip: Bootstrapping language-image pre-training for unified vision-language understanding and generation},
  author={Li, Junnan and Li, Dongxu and Xiong, Caiming and Hoi, Steven},
  booktitle={International conference on machine learning},
  pages={12888--12900},
  year={2022},
  organization={PMLR}
}

@inproceedings{zhai2023sigmoid,
  title={Sigmoid loss for language image pre-training},
  author={Zhai, Xiaohua and Mustafa, Basil and Kolesnikov, Alexander and Beyer, Lucas},
  booktitle={Proceedings of the IEEE/CVF international conference on computer vision},
  pages={11975--11986},
  year={2023}
}

@article{li2026qwen3,
  title={Qwen3-VL-Embedding and Qwen3-VL-Reranker: A Unified Framework for State-of-the-Art Multimodal Retrieval and Ranking},
  author={Li, Mingxin and Zhang, Yanzhao and Long, Dingkun and Chen, Keqin and Song, Sibo and Bai, Shuai and Yang, Zhibo and Xie, Pengjun and Yang, An and Liu, Dayiheng and others},
  journal={arXiv preprint arXiv:2601.04720},
  year={2026}
}

@article{zhang2024gme,
  title={GME: Improving Universal Multimodal Retrieval by Multimodal LLMs},
  author={Zhang, Xin and Zhang, Yanzhao and Xie, Wen and Li, Mingxin and Dai, Ziqi and Long, Dingkun and Xie, Pengjun and Zhang, Meishan and Li, Wenjie and Zhang, Min},
  journal={arXiv preprint arXiv:2412.16855},
  year={2024}
}

@article{jian2025rzenembed,
  title={Rzenembed: Towards comprehensive multimodal retrieval},
  author={Jian, Weijian and Zhang, Yajun and Liang, Dawei and Xie, Chunyu and He, Yixiao and Leng, Dawei and Yin, Yuhui},
  journal={arXiv preprint arXiv:2510.27350},
  year={2025}
}

@inproceedings{liu2025lamra,
  title={Lamra: Large multimodal model as your advanced retrieval assistant},
  author={Liu, Yikun and Zhang, Yajie and Cai, Jiayin and Jiang, Xiaolong and Hu, Yao and Yao, Jiangchao and Wang, Yanfeng and Xie, Weidi},
  booktitle={Proceedings of the Computer Vision and Pattern Recognition Conference},
  pages={4015--4025},
  year={2025}
}

@article{gu2025unime,
  title={Unime-v2: Mllm-as-a-judge for universal multimodal embedding learning},
  author={Gu, Tiancheng and Yang, Kaicheng and Zhang, Kaichen and An, Xiang and Feng, Ziyong and Zhang, Yueyi and Cai, Weidong and Deng, Jiankang and Bing, Lidong},
  journal={arXiv preprint arXiv:2510.13515},
  year={2025}
}

@inproceedings{zhang2025notellm,
  title={Notellm-2: Multimodal large representation models for recommendation},
  author={Zhang, Chao and Zhang, Haoxin and Wu, Shiwei and Wu, Di and Xu, Tong and Zhao, Xiangyu and Gao, Yan and Hu, Yao and Chen, Enhong},
  booktitle={Proceedings of the 31st ACM SIGKDD Conference on Knowledge Discovery and Data Mining V. 1},
  pages={2815--2826},
  year={2025}
}

@article{lee2024nv,
  title={Nv-embed: Improved techniques for training llms as generalist embedding models},
  author={Lee, Chankyu and Roy, Rajarshi and Xu, Mengyao and Raiman, Jonathan and Shoeybi, Mohammad and Catanzaro, Bryan and Ping, Wei},
  journal={arXiv preprint arXiv:2405.17428},
  year={2024}
}

@article{shao2024deepseekmath,
  title={Deepseekmath: Pushing the limits of mathematical reasoning in open language models},
  author={Shao, Zhihong and Wang, Peiyi and Zhu, Qihao and Xu, Runxin and Song, Junxiao and Bi, Xiao and Zhang, Haowei and Zhang, Mingchuan and Li, YK and Wu, Yang and others},
  journal={arXiv preprint arXiv:2402.03300},
  year={2024}
}

@article{liu2025taosearchemb,
  title={Taosearchemb: A multi-objective reinforcement learning framework for dense retrieval in taobao search},
  author={Liu, Xingxian and Li, Dongshuai and Wen, Tao and Wan, Jiahui and Ling, Gui and Lv, Fuyu and Ou, Dan and Tang, Haihong},
  journal={arXiv preprint arXiv:2511.13885},
  year={2025}
}

@article{jiang2024e5,
  title={E5-v: Universal embeddings with multimodal large language models},
  author={Jiang, Ting and Song, Minghui and Zhang, Zihan and Huang, Haizhen and Deng, Weiwei and Sun, Feng and Zhang, Qi and Wang, Deqing and Zhuang, Fuzhen},
  journal={arXiv preprint arXiv:2407.12580},
  year={2024}
}

@inproceedings{lin13microsoft,
  title={Microsoft coco: Common objects in context},
  author={Lin, Tsung-Yi and Maire, Michael and Belongie, Serge and Hays, James and Perona, Pietro and Ramanan, Deva and Doll{\'a}r, Piotr and Zitnick, C Lawrence},
  booktitle={European conference on computer vision},
  pages={740--755},
  year={2014},
  organization={Springer}
}

@inproceedings{liu2021visual,
  title={Visual news: Benchmark and challenges in news image captioning},
  author={Liu, Fuxiao and Wang, Yinghan and Wang, Tianlu and Ordonez, Vicente},
  booktitle={Proceedings of the 2021 conference on empirical methods in natural language processing},
  pages={6761--6771},
  year={2021}
}

@inproceedings{kazemzadeh2014referitgame,
  title={Referitgame: Referring to objects in photographs of natural scenes},
  author={Kazemzadeh, Sahar and Ordonez, Vicente and Matten, Mark and Berg, Tamara},
  booktitle={Proceedings of the 2014 conference on empirical methods in natural language processing (EMNLP)},
  pages={787--798},
  year={2014}
}

@inproceedings{deng2009imagenet,
  title={Imagenet: A large-scale hierarchical image database},
  author={Deng, Jia and Dong, Wei and Socher, Richard and Li, Li-Jia and Li, Kai and Fei-Fei, Li},
  booktitle={2009 IEEE conference on computer vision and pattern recognition},
  pages={248--255},
  year={2009},
  organization={Ieee}
}

@inproceedings{hendrycks2021many,
  title={The many faces of robustness: A critical analysis of out-of-distribution generalization},
  author={Hendrycks, Dan and Basart, Steven and Mu, Norman and Kadavath, Saurav and Wang, Frank and Dorundo, Evan and Desai, Rahul and Zhu, Tyler and Parajuli, Samyak and Guo, Mike and others},
  booktitle={Proceedings of the IEEE/CVF international conference on computer vision},
  pages={8340--8349},
  year={2021}
}

@inproceedings{hendrycks2021natural,
  title={Natural adversarial examples},
  author={Hendrycks, Dan and Zhao, Kevin and Basart, Steven and Steinhardt, Jacob and Song, Dawn},
  booktitle={Proceedings of the IEEE/CVF conference on computer vision and pattern recognition},
  pages={15262--15271},
  year={2021}
}

@article{zhou2017places,
  title={Places: A 10 million image database for scene recognition},
  author={Zhou, Bolei and Lapedriza, Agata and Khosla, Aditya and Oliva, Aude and Torralba, Antonio},
  journal={IEEE transactions on pattern analysis and machine intelligence},
  volume={40},
  number={6},
  pages={1452--1464},
  year={2017},
  publisher={IEEE}
}

@inproceedings{zhu2016visual7w,
  title={Visual7w: Grounded question answering in images},
  author={Zhu, Yuke and Groth, Oliver and Bernstein, Michael and Fei-Fei, Li},
  booktitle={Proceedings of the IEEE conference on computer vision and pattern recognition},
  pages={4995--5004},
  year={2016}
}

@article{everingham2015pascal,
  title={The pascal visual object classes challenge: A retrospective},
  author={Everingham, Mark and Eslami, SM Ali and Van Gool, Luc and Williams, Christopher KI and Winn, John and Zisserman, Andrew},
  journal={International journal of computer vision},
  volume={111},
  number={1},
  pages={98--136},
  year={2015},
  publisher={Springer}
}

@inproceedings{mathew2021docvqa,
  title={Docvqa: A dataset for vqa on document images},
  author={Mathew, Minesh and Karatzas, Dimosthenis and Jawahar, CV},
  booktitle={Proceedings of the IEEE/CVF winter conference on applications of computer vision},
  pages={2200--2209},
  year={2021}
}

@article{zheng2023judging,
  title={Judging llm-as-a-judge with mt-bench and chatbot arena},
  author={Zheng, Lianmin and Chiang, Wei-Lin and Sheng, Ying and Zhuang, Siyuan and Wu, Zhanghao and Zhuang, Yonghao and Lin, Zi and Li, Zhuohan and Li, Dacheng and Xing, Eric and others},
  journal={Advances in neural information processing systems},
  volume={36},
  pages={46595--46623},
  year={2023}
}

@inproceedings{chen2024mllm,
  title={Mllm-as-a-judge: Assessing multimodal llm-as-a-judge with vision-language benchmark},
  author={Chen, Dongping and Chen, Ruoxi and Zhang, Shilin and Wang, Yaochen and Liu, Yinuo and Zhou, Huichi and Zhang, Qihui and Wan, Yao and Zhou, Pan and Sun, Lichao},
  booktitle={Forty-first International Conference on Machine Learning},
  year={2024}
}

@article{yang2022chinese,
  title={Chinese clip: Contrastive vision-language pretraining in chinese},
  author={Yang, An and Pan, Junshu and Lin, Junyang and Men, Rui and Zhang, Yichang and Zhou, Jingren and Zhou, Chang},
  journal={arXiv preprint arXiv:2211.01335},
  year={2022}
}

\appendix

\section{Contrastive SFT Settings}

\begin{table}[!b]  
    \centering
    \caption{SFT training hyperparameters}
    \label{tab:SFT_setting}
    \begin{tabular}{ll}
        Hyperparameter & Value  \\ \hline
        Training samples & 900k  \\ 
        Batch size & 64  \\ 
        Learning rate & 1e-4  \\ 
        Warmup ratio & 0.1  \\ 
        LoRA rank & 16  \\ 
        Training steps & 4000  \\ 
        Optimizer & AdamW  \\ 
        Infra & GradCache  \\ 
        Max length & 4096  \\ 
        Temperature & 0.02  \\ 
        Hard negative num & 8  \\ 
        GPU config & $8 \times$ H800 \\
    \end{tabular}
\end{table}

\begin{table}[!h] 
    \centering
    \caption{RL training hyperparameters}
    \label{tab:RL_setting}
    \begin{tabular}{ll}
        Hyperparameter & Value  \\ \hline
        Training samples & 60k  \\ 
        Training epochs & 1  \\ 
        Batch size & 8  \\ 
        LoRA rank & 16  \\ 
        Learning rate & 5e-7  \\ 
        Precision & bf16 \\ 
        Penalty $(C_{\text{pen}})$ & -5 \\
        Base reward $(R_{\text{base}})$ & $3\times len(L_{rel})$ \\
    \end{tabular}
\end{table}

\begin{table}
    \centering
    \caption{Performance comparison with Chinese-CLIP (\%).}
    \label{tab:chinese_clip_comparison}
    \begin{tabular}{lccccc}
        \toprule
        \textbf{Method} & \textbf{I2T} & \textbf{T2I} & \textbf{I2OCR} & \textbf{OCR2I} & \textbf{Avg} \\
        \midrule
        Chinese-CLIP (\%) & 44.3 & 45.8 & 25.6 & 28.6 & 36.1 \\
        \bottomrule
    \end{tabular}
\end{table}

\begin{table}[ht]
    \centering
    \caption{Performance comparison between UniNote and Qwen3VL-Reranker-8B.}
    \label{tab:rerank_comparison}
    \begin{tabular}{lccc}
        \toprule
        \textbf{Model} & \textbf{Avg R@1} & \textbf{Avg R@5} & \textbf{Avg R@10} \\
        \midrule
        Qwen3VL-Reranker-8B & 0.79 & 0.86 & 0.90 \\
        UniNote  & 0.76 & 0.86 & 0.92 \\
        \bottomrule
    \end{tabular}
\end{table}

\begin{table}[ht]
    \centering
    \caption{Practical utility and training cost details.}
    \label{tab:training_costs}
    \begin{tabular}{l p{0.6\columnwidth}}
        \toprule
        \textbf{Item} & \textbf{Details} \\
        \midrule
        Training cost & 8x H800, 4 days (900k SFT / 60k RL samples) \\
        \midrule
        Storage \& index & MRL dimension: 64 / 512 / 1024 / 4096 \\
        \midrule
        Inference cost & Encode a note: 278 ms ($N_{image}=1$), 748 ms ($N_{image}=8$). \\
        \bottomrule
    \end{tabular}
\end{table}

We filter high-quality notes from the raw dataset of Xiaohongshu based on a threshold of more than 100 likes. Balanced processing is performed across topics and image counts per note to prevent weaknesses in specific dimensions. Hard negative samples are selected with $\tau_{\mathrm{min}}=0.5$ and $\tau_{\mathrm{max}}=0.7$, determined through manual sampling and inspection. This yields 900{,}000 training samples across nine task types (excluding note2note), each with 100{,}000 samples.

We follow UniME-V2~\cite{gu2025unime} training parameters to perform LoRA fine-tuning on Qwen3VL-8B-Instruct using eight H800 GPUs (80G). The specific parameters are listed in Table~\ref{tab:SFT_setting}.

\section{Relative RL Reranking Settings}
In this stage, we construct overlaps between notes by splitting them, producing 60k training samples. $C_{\text{pen}}$ and $R_{\text{base}}$ are hyperparameters in the reward function. The specific parameters are listed in Table ~\ref{tab:RL_setting}.

\section{Supplementary Experiments}

In this section, we provide additional experimental results to further validate the performance of UniNote, including comparisons with the Chinese-CLIP\cite{yang2022chinese} baseline and SOTA two-stage (retrieval-reranker) architectures. Before presenting the empirical results, we first clarify the evaluation metric unique to the Note2Note task.

\subsection{Evaluation Metrics and Protocols}
As noted in the main text, while standard tasks utilize the traditional $\mathrm{R}@\mathrm{K}$ metric, the Note2Note task involves multiple ground-truth candidates per query. To accommodate this, we employ the adjusted recall metric $\mathrm{R}^*@\mathrm{K}$, defined as follows:
\begin{equation}
\mathrm{R}^*@\mathrm{K} = \frac{1}{N}\sum_{q=1}^{N} \left( \frac{\sum_{x \in \mathrm{TopK}_q} \mathbb{I}(x \in R_q)}{|R_q|} \right)
\end{equation}
where $N$ is the total number of queries. For a specific query $q$, $R_q$ denotes the set of ground-truth relevant items, and  $\mathrm{TopK}_q$) represents the top-$K$ retrieved candidates. On our evaluation test set, the average size of the relevant set is $\mathrm{Avg}(|R_q|) = 6.68$, and the theoretical maximum upper bound for $\mathrm{R}^*@1$ is $16.2\%$.
\subsection{Comparison with Chinese-CLIP}
As presented in Table~\ref{tab:chinese_clip_comparison}, Chinese-CLIP achieves an average $\mathrm{R}@1$ of 36.1\% on a 10k-sized gallery. While established CLIP-based models perform well in general-purpose retrieval, a distinct performance gap persists compared to our MLLM-based approach. We attribute this difference to the inherent limitations of CLIP-based models regarding context window constraints and embedding granularity. Conversely, UniNote leverages a more robust visual-textual alignment, which facilitates superior fine-grained discrimination and semantic representation, particularly in large-scale retrieval scenarios.

\subsection{Comparison with Two-Stage Models}
To evaluate the trade-off between precision and efficiency, we compared UniNote against a SOTA reranker (Qwen3VL-Reranker-8B) using a small-scale demo setup. As summarized in Table~\ref{tab:rerank_comparison}, although a marginal performance gap remains in R@1, UniNote achieves performance comparable to the complex two-stage paradigm while operating within a single embedding pass. By eliminating the high-latency reranking stage, our model satisfies the stringent latency and computational cost constraints typical of real-world production environments. Future work will focus on further narrowing the performance gap by scaling the training data and enhancing our hierarchical reward mechanisms.

\section{Practical Utility and Training Costs}

We evaluate the practical utility of our embedding-based retrieval system through the lens of computational efficiency and deployment feasibility.  Table~\ref{tab:training_costs} provides a detailed breakdown of the resource requirements and performance metrics associated with our implementation.

Furthermore, the MLLM backbone offers inherent flexibility, allowing for dynamic adjustments to the input image sequence. This capability enables practitioners to manage computational budgets effectively by balancing retrieval precision against latency requirements based on specific application constraints.

\end{document}